# Working Paper

# A Non Parametric Model for the Forecasting of Venezuelan Oil Prices

Sabatino Costanzo, Loren Trigo, Wafaa El Dehne, Hender Prato

# Papel de Trabajo

# Modelo No Paramétrico para la Predicción del Movimiento de los Precios del Crudo Venezolano

Sabatino Costanzo, Loren Trigo, Wafaa El Dehne, Hender Prato

**Abstract**

A neural net model for forecasting the prices of Venezuelan crude oil is proposed. The inputs of the neural net are selected by reference to a dynamic system model of oil prices by Mashayekhi (1995, 2001) and its performance is evaluated using two criteria: the Excess Profitability test by Anatoliev and Gerko (2005) and the characteristics of the equity curve generated by a trading strategy based on the neural net's predictions.

**Resumen**

Se introduce aquí un modelo no paramétrico para pronosticar los precios del petróleo Venezolano cuyos insumos son seleccionados en base a un sistema dinámico que explica los precios en términos de dichos insumos. Se describe el proceso de recolección y pre-procesamiento de datos y la corrida de la red y se evalúan sus pronósticos a través de un test estadístico de predictibilidad y de las características del *Equity Curve* inducido por la estrategia de compraventa bursátil generada por dichos pronósticos.

# Introducción

A lo largo de las últimas décadas han tenido lugar innumerables intentos de desarrollar modelos que permitan pronosticar el movimiento de los precios de los commodities en el mercado, reduciendo la incertidumbre durante el proceso de compra-venta bursátil. Entre las herramientas utilizadas se encuentran los métodos no paramétricos (Kernel Regressions, Support Vector Machines, Redes Neurales) y los Sistemas Dinámicos, pero utilizados en forma independiente. Por tal razón se plantean las preguntas naturales: ¿Existe alguna manera de conjugar el uso de estas herramientas para lograr predecir los precios con mayor exactitud que utilizando cualquiera de ellos por separado? ¿Permite la integración de los métodos regresivos no-paramétricos a los sistemas dinámicos generar modelos que predigan con mayor exactitud el precio del petróleo que estas mismas herramientas utilizadas en forma independiente? De estas preguntas se derivan otras: Entre todas las posibles, ¿Cuál es la forma más eficiente de insertar estos métodos no-paramétricos en los sistemas dinámicos? ¿Es posible GENERALIZAR la respuesta a la pregunta central, es decir, mejorar sensiblemente los resultados obtenidos de los métodos regresivos no-paramétricos no solo en el caso del petróleo sino EN GENERAL, sistematizando racionalmente la selección de *inputs* utilizados para alimentarlos a través del uso de sistemas dinámicos?

La utilización de Sistemas Dinámicos en la predicción de precios se remonta a más de una década atrás. La utilización de métodos no paramétricos con fines predictivos (Recopilación de ensayos y artículos de Murray Ruggieiro: Futures Magazine, 1990-2003) anticipa la integración de ambas metodologías (Alborzi [1]: "Implanting Neural Network Elements in System Dynamics Models to Surrogate Rate and Auxiliary Variables) y promete ser una fuente de nuevas iniciativas para lograr resultados importantes.

En este Papel de Trabajo, tras la recopilación de la bibliografía relevante en el área de sistemas dinámicos aplicada al proceso de formación del precio de los commodities en general y la adopción del uso del *software* IThink & Vensym para articular la teoría de formación de precios de *commodities* propuesta por J. Forrester en *Business Dynamics*.

nálisis de las diferentes alternativas de inserción de métodos no paramétricos en sistemas dinámicos disponibles en la literatura, se procedió a la conformación de una cesta de crudos de composición análoga a la de la cesta petrolera de la OPEC, para proceder a la determinación y recopilación de la data requerida para alimentar *los nodos* del sistema dinámico, a ser utilizados como inputs que inciden en el precio de los componentes de la cesta de crudo. Luego de la generación de señales de compra-venta de los distintos tipos de crudo en la cesta representativa, se procedió a efectuar un *testing* histórico del sistema en donde se le requirió el pronóstico del precio diario de la cesta de crudos sobre 10 años de data diaria. Se evaluó el portafolio en términos de los *benchmarks* Mínimo & Máximo *(Buy and hold & Perfect equity)* obteniendo como resultado un modelo que permite pronosticar el precio de una cesta de crudo diversificada con un nivel de error satisfactorio.

# Modelo Predictivo

La naturaleza del problema que es objeto de este estudio en el cual las relaciones entre las variables no son lineales, plantea la necesidad de recurrir a un modelo no paramétrico basado en redes neurales.

**Redes Neurales**

No existe un método formal para la selección de las mismas y una mala elección puede producir un pobre desempeño (Walczak, 1999).

En este caso, si bien es claro que la variable dependiente del modelo es el precio de la cesta petrolera venezolana, al mismo tiempo surge la interrogante de cuáles variables de entrada son las mas adecuadas para predecir el resultado con un márgen de error razonable. A fin de responder a esta pregunta de acuerdo a los principios del *Intermarket Analysis*, a veces se recurre a la evidencia empírica de que el precio de ciertos bienes está correlacionado con el comportamiento de otros (Ruggiero, 2003: correlación negativa de los T-Bonds con precio del crudo; correlación positiva del precio del crudo con el XAU Index; correlación negativa del precio del crudo vs. USD Index, etc.). Así mismo se recurre a la selección de estas variables de entrada mediante la heurística, o bien al empleo de modelos ya validados y probados. En este trabajo se utilizará un modelo validado consistente en un sistema dinámico[1] que relaciona el precio del petróleo con varias variables macroeconómicas relevantes. Este modelo, desarrollado por Ali N. Mashayekhi y publicado en su artículo *Dynamics of Oil Price in the World Market*, 2001, sugiere que los precios del petróleo presentan un comportamiento oscilatorio e incluye entre sus variables la Infra-Estructura Dependiente de Ingresos Petroleros, la Infra-Estructura en Desarrollo, el Precio del Petróleo, la Demanda Base de Petróleo, la Reserva de Ingresos Petroleros y las Reservas Petroleras Probadas.

Según el autor, cuando el precio del petróleo aumenta, hay una reacción tanto de los países productores como de los consumidores, en donde por un lado, los primeros incrementan sus recursos económicos en el corto plazo, por lo cual sus necesidades para exportar petróleo caen. En consecuencia de esto, el suministro decrece, conllevando en el mediano plazo, al incremento del precio. Por otra parte, los países consumidores reaccionan disminuyendo la demanda base con cierto rezago en el tiempo. De esta forma, mientras los países consumidores disminuyen la demanda, los exportadores gastan sus ingresos crecientes provenientes del petróleo en el desarrollo una estructura financiera gubernamental cada vez más dependiente de este tipo de ingresos, lo cual conlleva al aumento creciente de sus necesidades de exportar. Así, un suministro creciente aunado a la disminución de la demanda de los países consumidores trae como consecuencia un colapso que se manifiesta en la caída abrupta de los precios. Precios cada vez más bajos implican el descenso de los ingresos de los países exportadores, los cuales, debido a sus necesidades de ingresos cada vez mayores, tratan de compensar el déficit incrementando sus exportaciones, acentuando aún más con el exceso de oferta, la caída de los precios. Todo esto da origen a un movimiento reactivo que, por una parte comienza incipientemente a hacer disminuir la dependencia de los productores por este tipo de ingresos, al mismo tiempo que hace resurgir incipientemente una nueva espiral ascendente de demanda de petróleo por parte de los países consumidores. Así, la dependencia decreciente de los ingresos petroleros de los productores que disminuye la presión que estos sienten a exportar, al coexistir con un incremento progresivo de la demanda por parte de los consumidores, ejerce una nueva presión ascendente sobre el precio del petróleo, dando así origen a un nuevo ciclo.

Al profundizar sobre el tipo de información que alimenta las variables contempladas como insumos del modelo, el autor define la estructura dependiente de los ingresos petroleros como conformada por la capacidad de producción creada tanto en el sector público como en el privado. Salud, educación, seguridad nacional, servicios públicos e infraestructura de producción, son, según él, los rubros centrales en los cuales se emplean los gastos nacionales destinados al consumo, por lo cual se asocia esta variable al consumo total público y privado extraído de las cuentas consolidadas de la nación. La variable Estructura en Desarrollo se interpreta como la formación bruta de capital fijo, referido a la parte del gasto nacional destinado a la inversión. Respecto al resto de las variables, el Precio del Petróleo corresponde al precio de cierre semanal para el crudo venezolano, el cual servirá como insumo del modelo junto a la Demanda Base de este tipo de crudo, a la Reserva de Ingresos Derivados de las Exportaciones Petroleras y a las Reservas Petroleras per-se. El modelo supone a esta última como una variable con gran número de reservas iniciales, por lo tanto no limitará las exportaciones por un largo período de tiempo y no influirá, por el momento, en el comportamiento oscilatorio del precio.

En nuestro estudio se desarrollan dos modelos de predicción para pronosticar el precio de la cesta petrolera venezolana. El primero estará alimentado con información exclusiva de Venezuela como primera aproximación. Como resulta difícil y contraintuitivo pensar que las decisiones aisladas de un país son las únicas que afectan los precios de sus productos en el exterior –es el mercado mundial, como un agregado, quien determina el valor– , se desarrollará un segundo modelo de predicción que se alimentará con el mismo tipo de información recabada para Venezuela para otros países exportadores de petróleo miembros de la OPEP como Arabia Saudita, Irán y Kuwait. En el estudio *Government Financial Structure and Oil Revenues Expenditures in Iran* (Mashayeki, 1995), el autor presenta la analogía de los comportamientos del crecimiento de los ingresos nacionales en estos cuatro países a partir del desarrollo del mercado petrolero en la década de los 70. Curiosamente, tanto en el crecimiento como en el margen de dependencia de estos ingresos con respecto a la renta petrolera, se observa, con escasa diferencia, la similitud entre el caso venezolano y el iraní.

Tras generar la red neural, entrenarla sobre los inputs y outputs discutidos y obtener los resultados predictivos del modelo, se valida su rentabilidad diseñando y evaluando una estrategia de compraventa bursátil en el mercado de contratos a futuro de petróleo que consiste en entrar al mercado con una posición acorde al pronóstico del alza o la baja del precio del petróleo midiendo el rendimiento de las operaciones bajadas en dicha estrategia.

## Metodología

**Recolección de la información**

Las variables descritas fueron observadas semanalmente durante el período 1997-2005.

Los datos utilizados fueron tomados de las páginas Web: *Organization of Petroleum Exporting Countries* (OPEC), *Internacional Energy Agency* (IEA), Banco Central de Venezuela (BCV), así como del anuario *Internacional Financial Statistics* (IFS) de las Naciones Unidas.

Los datos semanales, trimestrales y anuales requirieron para la homologación, de un pre-procesamiento previo a la corrida de la red que consistió en una serie de interpolaciones

**Procesamiento**

Los precios del petróleo se obtuvieron en base semanal; los ingresos, en base trimestral y los gastos, inversión y demanda, en base anual. Para predecir los precios del petróleo se homologó toda la data a la unidad de tiempo semanal a través de una interpolación polinómica con un factor de correlación $R^2$ cercano a 1.

A continuación se procedió al suavizamiento *(Smoothing)* para facilitar el proceso de aprendizaje de la red mediante el uso de promedios móviles sencillos para todos los *inputs*, uno corto con el promedio de los últimos dos datos ocurridos y uno largo con el de los últimos cinco (ver figura 1).

Para el precio del petróleo, el *output* de la red, se calculó un promedio móvil de cuatro datos proyectado cinco semanas en el futuro para predecir el precio en la quinta semana (ver figura 1). Los datos suavizados forman parte de los *inputs* de la red.

Figura 1. Promedios móviles de dos y de cinco para los *Inputs* y promedio móvil de cuatro para el *output*

| Año | N° datos semanal | Gasto Consumo Total (millones Bs) | mav 2 | mav 5 | lag 15 | Ingresos petroleros (millones $) | mav 7 | mav 15 | lag 19 | Demanda (1000 b/d) | mav 8 | mav 16 | lag 0 | Inversión (millones Bs) | mav 2 | mav 5 | lag 15 | Precio petróleo($/b) | mav 4 |
|---|---|---|---|---|---|---|---|---|---|---|---|---|---|---|---|---|---|---|---|
| 1.997 | 1 | 10.396.381 | | | | 2.000 | | | | 1980,9 | | | 1980,9 | 5.185.167 | | | | 26,62 | |
| | 2 | 10.789.372 | | | | 2.126 | | | | 1985,4 | | | 1985,4 | 5.368.318 | | | | 26,62 | |
| | 3 | 11.178.994 | 10.984.183 | | | 2.253 | 2.190 | | | 1989,9 | 1.988 | | 1989,9 | 5.549.463 | 5.458.890 | | | 26,62 | |
| | 4 | 11.565.267 | 11.372.130 | | | 2.379 | 2.316 | | | 1994,4 | 1.992 | | 1994,4 | 5.728.613 | 5.639.038 | | | 26,62 | |
| | 5 | 11.948.212 | 11.756.739 | | | 2.506 | 2.442 | | | 1998,9 | 1.997 | | 1998,9 | 5.905.781 | 5.817.197 | | | 22,37 | 26,62 |
| | 6 | 12.327.849 | 12.138.030 | 11.175.645 | | 2.632 | 2.569 | 2.253 | | 2003,4 | 2.001 | 1.999 | 2003,4 | 6.080.976 | 5.993.379 | 5.547.468 | | 22,37 | 25,56 |
| | 7 | 12.704.199 | 12.516.024 | 11.561.939 | | 2.758 | 2.695 | 2.379 | | 2008,0 | 2.006 | 2.003 | 2008,0 | 6.254.211 | 6.167.594 | 5.726.630 | | 22,37 | 24,50 |
| | 8 | 13.077.283 | 12.890.741 | 11.944.904 | | 2.885 | 2.822 | 2.506 | | 2012,5 | 2.010 | 2.008 | 2012,5 | 6.425.495 | 6.339.853 | 5.903.809 | | 22,37 | 23,43 |
| | 9 | 13.447.122 | 13.262.202 | 12.324.562 | | 3.011 | 2.948 | 2.632 | | 2017,0 | 2.015 | 2.012 | 2017,0 | 6.594.841 | 6.510.168 | 6.079.015 | | 20,37 | 22,37 |
| | 10 | 13.813.735 | 13.630.428 | 12.700.933 | | 3.137 | 3.074 | 2.758 | | 2021,5 | 2.019 | 2.017 | 2021,5 | 6.762.258 | 6.678.549 | 6.252.261 | | 20,37 | 21,87 |
| | 11 | 14.177.144 | 13.995.440 | 13.074.038 | | 3.264 | 3.201 | 2.885 | | 2026,0 | 2.024 | 2.021 | 2026,0 | 6.927.759 | 6.845.009 | 6.423.556 | | 20,37 | 21,37 |
| | 12 | 14.537.369 | 14.357.257 | 13.443.897 | | 3.390 | 3.327 | 3.011 | | 2030,5 | 2.028 | 2.026 | 2030,5 | 7.091.355 | 7.009.557 | 6.592.913 | | 20,37 | 20,87 |
| | 13 | 14.894.432 | 14.715.900 | 13.810.531 | | 3.517 | 3.453 | 3.137 | | 2035,0 | 2.033 | 2.030 | 2035,0 | 7.253.056 | 7.172.205 | 6.760.342 | | 20,37 | 20,37 |
| | 14 | 15.248.351 | 15.071.391 | 14.173.960 | | 3.677 | 3.597 | 3.264 | | 2039,5 | 2.037 | 2.035 | 2039,5 | 7.412.873 | 7.332.964 | 6.925.854 | | 20,37 | 20,37 |
| | 15 | 15.599.149 | 15.423.750 | 14.534.206 | | 3.711 | 3.694 | 3.397 | | 2044,0 | 2.042 | 2.040 | 2044,0 | 7.570.818 | 7.491.845 | 7.089.460 | | 20,37 | 20,37 |
| | 16 | 15.946.846 | 15.772.997 | 14.891.289 | 10.396.381 | 3.744 | 3.727 | 3.512 | | 2048,5 | 2.046 | 1.886,7 | 2048,5 | 7.726.902 | 7.648.860 | 7.251.172 | 5.185.167 | 16,32 | 20,37 |
| | 17 | 16.291.462 | 16.119.154 | 15.245.229 | 10.789.372 | 3.778 | 3.761 | 3.608 | | 2053,0 | 2.051 | 2.014,7 | 2053,0 | 7.881.135 | 7.804.019 | 7.411.001 | 5.368.318 | 16,32 | 19,36 |
| | 18 | 16.633.017 | 16.462.240 | 15.596.048 | 11.178.994 | 3.812 | 3.795 | 3.685 | | 2057,5 | 2.055 | 2.019,2 | 2057,5 | 8.033.530 | 7.957.333 | 7.568.957 | 5.549.463 | 16,32 | 18,35 |
| | 19 | 16.971.534 | 16.802.276 | 15.943.765 | 11.565.267 | 3.846 | 3.829 | 3.744 | | 2062,0 | 2.060 | 2.023,7 | 2062,0 | 8.184.097 | 8.108.813 | 7.725.052 | 5.728.613 | 16,32 | 17,33 |
| | 20 | 17.307.032 | 17.139.283 | 16.288.402 | 11.948.212 | 3.879 | 3.863 | 3.778 | 2.000 | 2066,6 | 2.064 | 2028,2 | 2066,6 | 8.332.846 | 8.258.472 | 7.879.296 | 5.905.781 | 16,32 | 16,32 |
| | 21 | 17.639.532 | 17.473.282 | 16.629.978 | 12.327.849 | 3.913 | 3.896 | 3.812 | 2.126 | 2071,1 | 2.069 | 2032,7 | 2071,1 | 8.479.790 | 8.406.318 | 8.031.702 | 6.080.976 | 18,22 | 16,32 |
| | 22 | 17.969.054 | 17.804.293 | 16.968.515 | 12.704.199 | 3.947 | 3.930 | 3.846 | 2.253 | 2075,6 | 2.073 | 2037,3 | 2075,6 | 8.624.939 | 8.552.365 | 8.182.280 | 6.254.211 | 18,22 | 16,80 |
| | 23 | 18.295.620 | 18.132.337 | 17.304.034 | 13.077.283 | 3.981 | 3.964 | 3.879 | 2.379 | 2080,1 | 2.078 | 2041,8 | 2080,1 | 8.768.305 | 8.696.622 | 8.331.041 | 6.425.495 | 19,02 | 17,27 |
| | 24 | 18.619.249 | 18.457.434 | 17.636.554 | 13.447.122 | 4.014 | 3.998 | 3.913 | 2.506 | 2084,6 | 2.082 | 2046,3 | 2084,6 | 8.909.898 | 8.839.102 | 8.477.996 | 6.594.841 | 19,02 | 17,95 |
| | 25 | 18.939.963 | 18.779.606 | 17.966.097 | 13.813.735 | 4.048 | 4.031 | 3.947 | 2.632 | 2089,1 | 2.087 | 2050,8 | 2089,1 | 9.049.730 | 8.979.814 | 8.623.156 | 6.762.258 | 18,62 | 18,62 |
| | 26 | 19.257.781 | 19.098.872 | 18.292.608 | 14.177.144 | 4.082 | 4.065 | 3.981 | 2.758 | 2093,6 | 2.091 | 2055,3 | 2093,6 | 9.187.811 | 9.118.770 | 8.766.533 | 6.927.759 | 19,02 | 18,82 |
| | 27 | 19.572.726 | 19.415.253 | 18.616.333 | 14.537.369 | 4.207 | 4.144 | 4.014 | 2.885 | 2098,1 | 2.096 | 2059,8 | 2098,1 | 9.324.153 | 9.255.982 | 8.908.137 | 7.091.355 | 19,02 | 19,02 |
| | 28 | 19.884.816 | 19.728.771 | 18.937.068 | 14.894.432 | 4.331 | 4.269 | 4.066 | 3.011 | 2102,6 | 2.100 | 2064,3 | 2102,6 | 9.458.767 | 9.391.460 | 9.047.979 | 7.253.056 | 19,02 | 19,02 |
| | 29 | 20.194.074 | 20.039.445 | 19.254.907 | 15.248.351 | 4.456 | 4.394 | 4.137 | 3.137 | 2107,1 | 2.105 | 2068,8 | 2107,1 | 9.591.663 | 9.525.215 | 9.186.072 | 7.412.873 | 19,02 | 19,02 |
| | 30 | 20.500.519 | 20.347.296 | 19.569.872 | 15.599.149 | 4.580 | 4.518 | 4.225 | 3.264 | 2111,6 | 2.109 | 2073,3 | 2111,6 | 9.722.854 | 9.657.259 | 9.322.425 | 7.570.818 | 19,02 | 19,02 |
| | 31 | 20.804.172 | 20.652.346 | 19.881.983 | 15.946.846 | 4.705 | 4.643 | 4.331 | 3.390 | 2116,1 | 2.114 | 2077,8 | 2116,1 | 9.852.350 | 9.787.602 | 9.457.050 | 7.726.902 | 19,02 | 19,02 |
| | 32 | 21.105.054 | 20.954.613 | 20.191.261 | 16.291.462 | 4.830 | 4.767 | 4.456 | 3.517 | 2120,6 | 2.118 | 2082,3 | 2120,6 | 9.980.162 | 9.916.256 | 9.589.957 | 7.881.135 | 19,02 | 19,02 |
| | 33 | 21.403.186 | 21.254.120 | 20.497.727 | 16.633.017 | 4.954 | 4.892 | 4.580 | 3.677 | 2125,2 | 2.123 | 2086,8 | 2125,2 | 10.106.301 | 10.043.232 | 9.721.159 | 8.033.530 | 19,02 | 19,02 |
| | 34 | 21.698.588 | 21.550.887 | 20.801.401 | 16.971.534 | 5.079 | 5.017 | 4.705 | 3.711 | 2129,7 | 2.127 | 2091,3 | 2129,7 | 10.230.779 | 10.168.540 | 9.850.666 | 8.184.097 | 19,02 | 19,02 |
| | 35 | 21.991.280 | 21.844.934 | 21.102.304 | 17.307.032 | 5.204 | 5.141 | 4.830 | 3.744 | 2134,2 | 2.132 | 2095,9 | 2134,2 | 10.353.606 | 10.292.192 | 9.978.489 | 8.332.846 | 19,02 | 19,02 |
| | 36 | 22.281.284 | 22.136.282 | 21.400.456 | 17.639.532 | 5.328 | 5.266 | 4.954 | 3.778 | 2138,7 | 2.136 | 2100,4 | 2138,7 | 10.474.793 | 10.414.200 | 10.104.640 | 8.479.790 | 19,02 | 19,02 |

La introducción de rezagos (*lags*) entre los insumos de la red está fundamentada en el hecho de que muchos de los efectos que ocurren en el ámbito macroeconómico y comercial no suceden inmediatamente a sus causas. Los precios del petróleo reaccionan con retardo respecto a la variación de los factores que sobre ellos influyen. Para lograr una buena estimación del *output* es necesario entonces hallar rezagos relevantes en el sentido financiero a través del test de predictibilidad de Anatolyev y Gerko[2].

Figura 2. Inserción de los *lags* para todos los *inputs*

Figura 3. Corte superior

**Corrida de la red neural**

Para la creación de la red neural se utilizó la aplicación Braincel®.

El primer paso es la distribución y definición de los datos de tal manera que el 60% sea considerada como rango de entrenamiento (*train*), el siguiente 30% como rango de prueba (*test*) y el 10% restante como datos nuevos (*new data*).

Un panorama general de la definición de los rangos se puede observar en la figura 4. Cabe observar que sólo se debe incluir el *output* observado en la corrida de entrenamiento.

Primero, la red debe reconocer y entender las relaciones entre las variables de entrada y la de salida para ser capaz de determinar las reglas a las cuales responde el patrón de comportamiento de los precios; segundo, evaluar la ejecución de la red en el rango de prueba contraponiendo su *output* calculado a los datos históricos observados y por último, observar el desempeño sobre los datos nuevos. La evaluación del rango de prueba sirve para preveer el desempeño de la red sobre los datos nuevos.

Figura 4. Vista completa de la definición de los rango

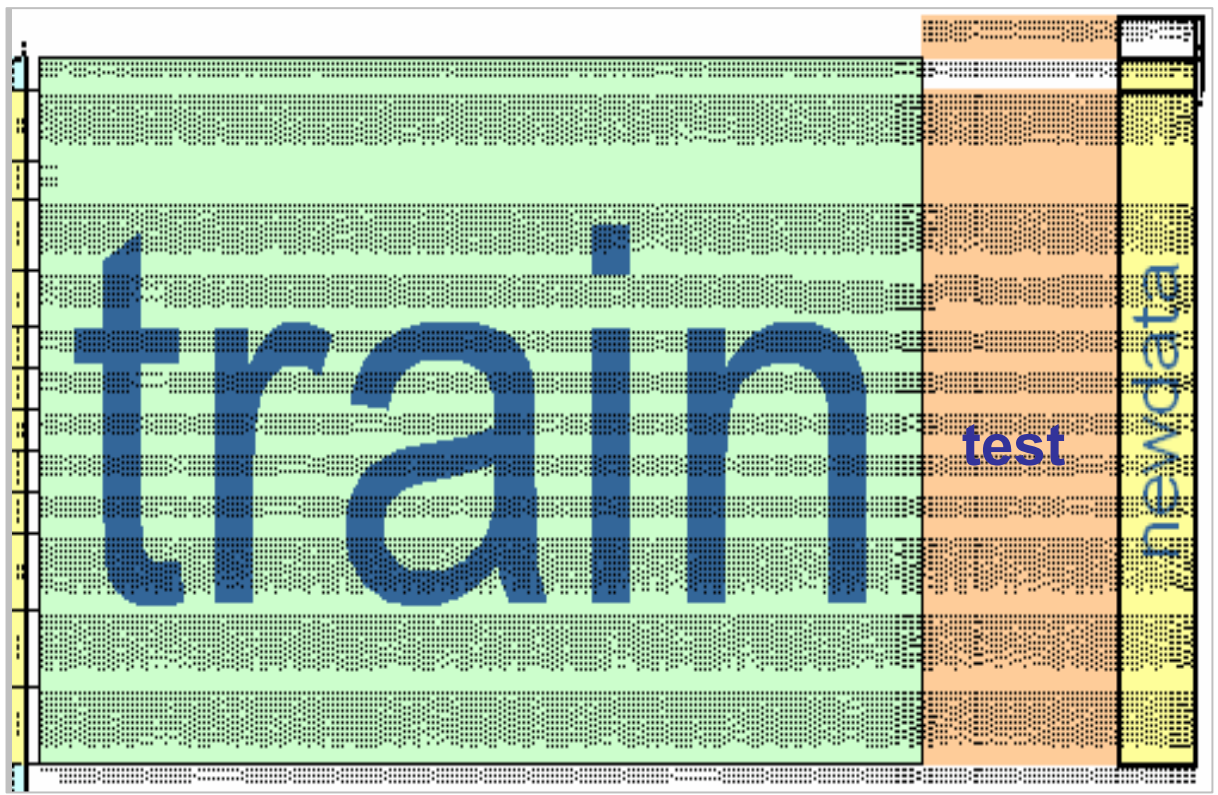

Una vez clasificados y definidos los datos, se crea una serie de expertos, los cuales deben ser guardados para efectos de comparación sobre el *Train Range* a fin de seleccionar el mas apto.

En el caso particular de este estudio se inició el entrenamiento de la red sobre el *Train Range* con los siguientes parámetros:

| | |
|---|---|
| Margen de error: | 5% |
| Tasa de aprendizaje: | 0,3 |
| Rango Peso inicial: | 0,4 |

hasta obtener convergencia.

Una vez entrenada, *cada* red generada fue probada y evaluada en el *Test Range*. Este procedimiento se repitió hasta obtener una red que lograra una buena estimación del precio según el test de Anatolyev y Gerko (ver gráfico 4). Una vez grabada la red que obtuvo la mejor evaluación, esta se corrió sobre el *New Data Range*, obteniendo el resultado que presentamos en este trabajo.

Grafico 4. Corrida de la red

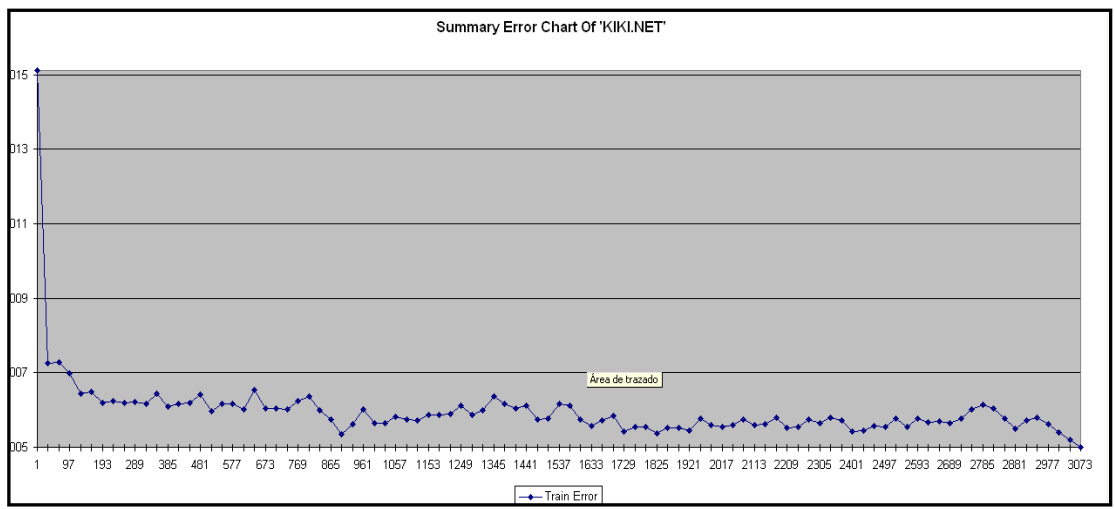

**Aplicación del Test de Anatolyev y Gerko**

Para evaluar la calidad del predictor se aplica el test de Anatolyev y Gerko. Esta técnica, comparado con otros (*Sharpe Ratio, Pesaran-Timmernann Test*), ha demostrado un óptimo desempeño evaluando la dirección y magnitud de la estimación.

En este proceso se tomaron las columnas de ambos *output* –calculado y observado– en el mismo orden y se sometieron a la hoja de cálculo de Anatoliev y Gerko. Al final de la prueba se obtuvo una probabilidad de predicción de 0,82% por lo que se procedió a correr la red sobre el rango *new data*. En esta ocasión se alcanzó un índice mayor a 0,99% (ver figura 5)

Figura 5. Corrida del rango de nuevos datos y su validación según test de Anatoliev y Gerko.

Los valores observados y calculados son cotejados entre sí, así como los valores arrojados por la red respecto a los valores reales (ver gráfico 5).

La curva resultante no muestra el comportamiento oscilatorio de los precios reales, sin embargo, refleja la misma tendencia creciente. La eficiencia predictiva de este modelo se verá claramente cuando se defina la estrategia de compraventa bursátil.

Grafico 5. Resultados de la red neural: caso Venezuela

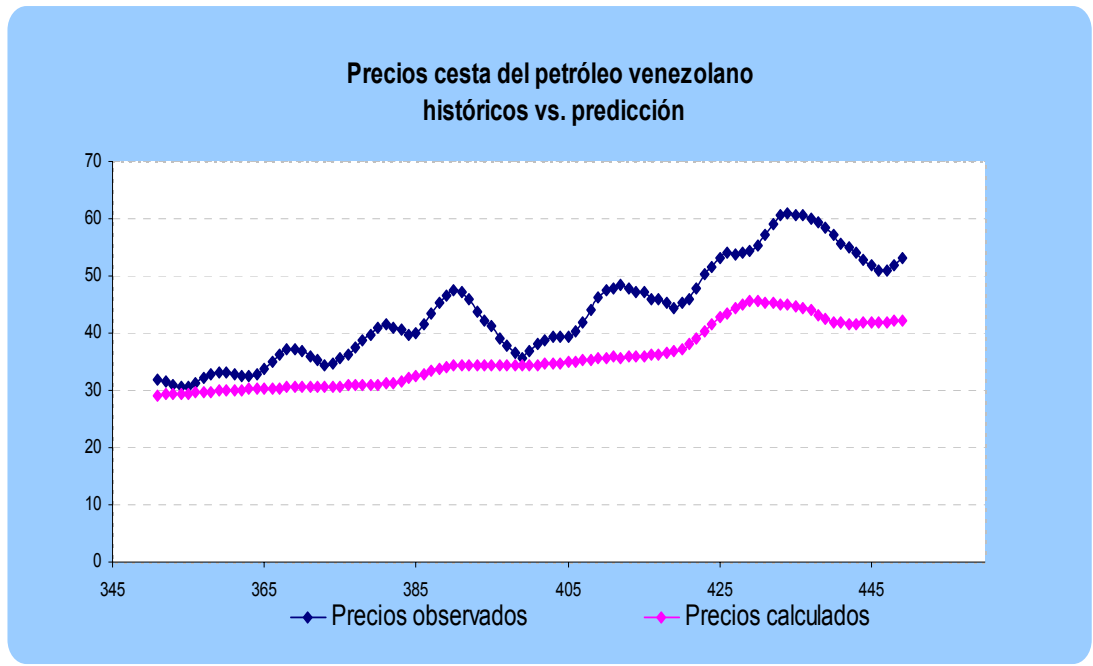

**Corrida de la red neural con información de cuatro países exportadores OPEP**

En el caso anterior se alimentó la red con inputs relacionados directamente con el caso de Venezuela. A continuación se muestra el resultado de realizar otra corrida incluyendo los datos de otros tres países también pertenecientes a la OPEP, con variables equivalentes a las venezolanas. El procesamiento de los datos se realizó de la misma forma que en el caso anterior (ver gráfico 6). En este nuevo caso el test de Anatoliev y Gerko confirmó con una alta probabilidad la viabilidad de la red seleccionada, tanto para el rango de prueba como para el de datos nuevos.

Grafico 6. Resultados de la red neural: caso Venezuela, Arabia Saudita, Irán y Kuwait

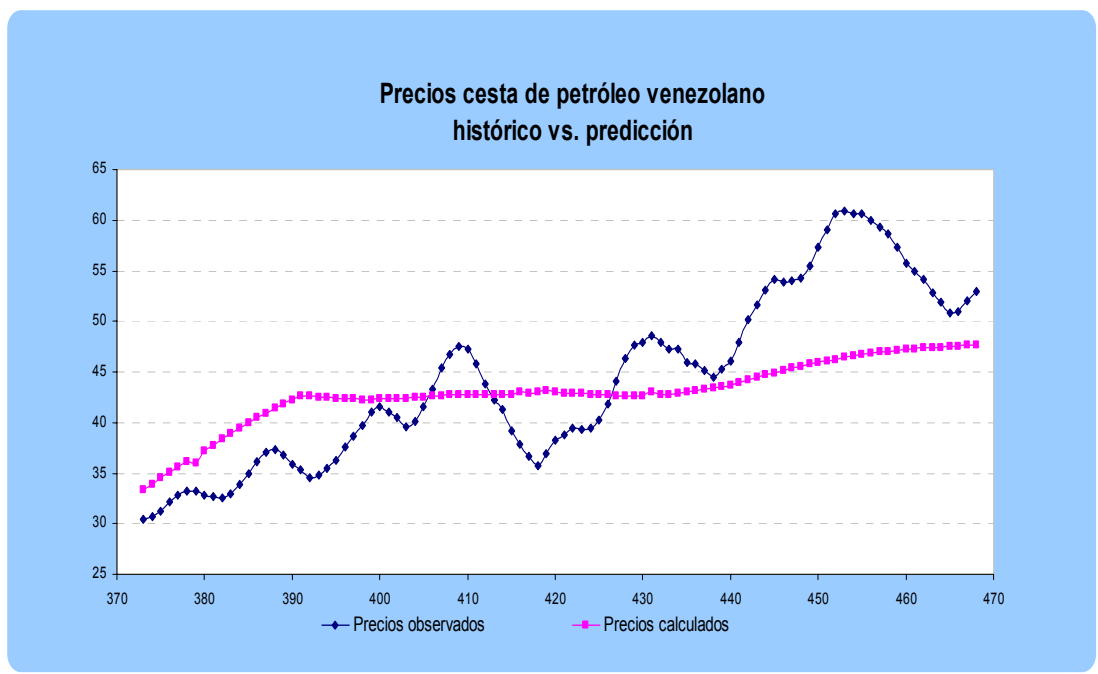

**Equity curve de los modelos predictivos**

Dado que el modelo de los cuatro países muestra un indicador de Anatolyev máximo, se procedió al cálculo del *equity curve* para medir la rentabilidad de la inversión basada en las predicciones del modelo. Sin embargo, se pudo observar que la ganancia sobre el capital invertido no era significativa al considerar los costos de transacción, lo cual muestra el hecho de que el exámen del *equity curve* inducido por las decisiones de mercado tomadas en base a las recomendaciones del experto virtual es mas refinado que el test de predictibilidad de Anatolyev, el cual, según muestra este ejemplo, es solo condición necesaria --mas no suficiente-- para determinar la rentabilidad de un experto virtual.

Es por eso que a partir de este momento se utilizó el *equity curve* para juzgar y seleccionar la red que habríamos de utilizar en el *New Data Range*.

Se procedió a la búsqueda de una red que reflejara una estrategia confiable y rentable para el inversionista realizando corridas sucesivas de redes generadas mediante cambios aleatorios de los coeficientes que determinan la red hasta obtener una satisfactoria (gráficos 7 & 8).

Grafico 7. Modelo mejorado que incluye inputs de los cuatro países

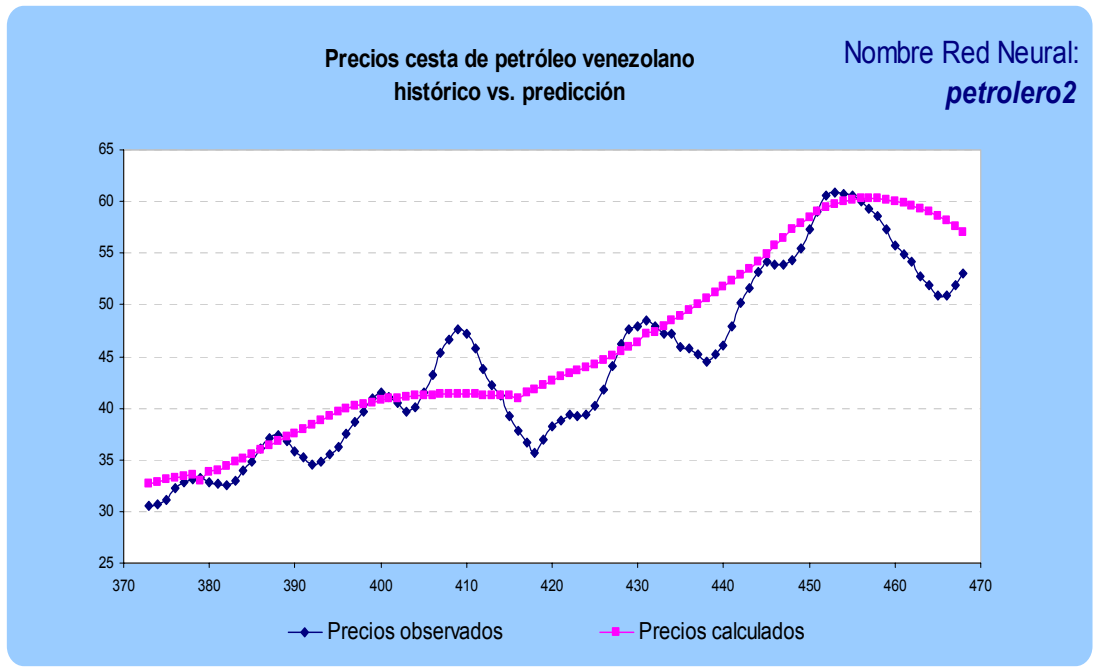

Grafico 8. *Equity curve* del modelo mejorado

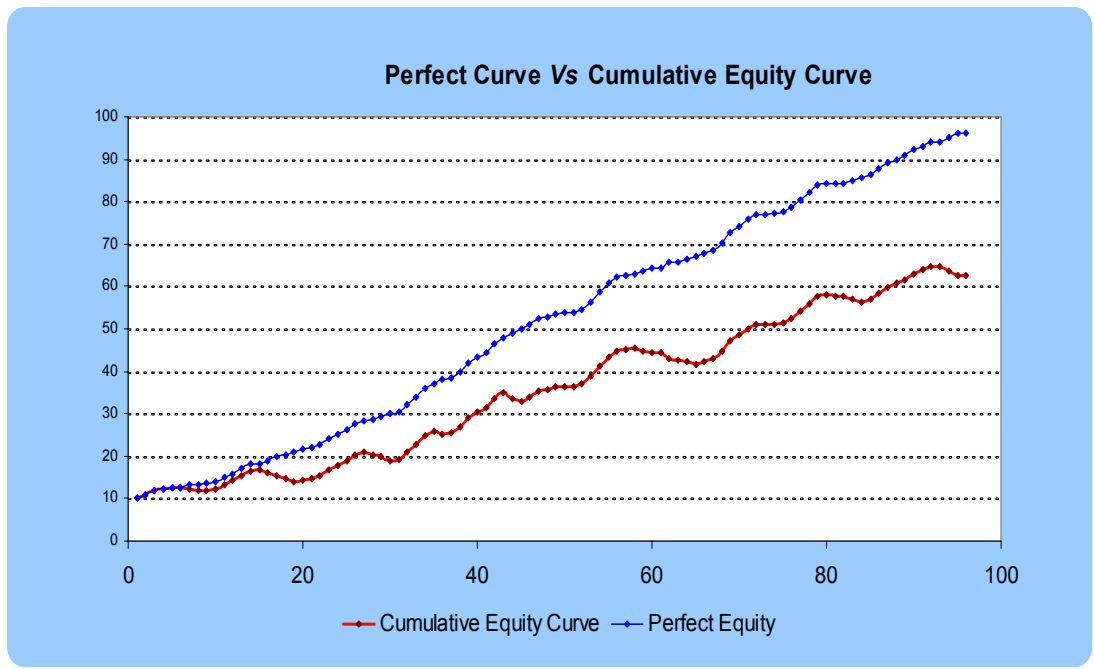

---

[4] *Perfect Equity*: es la ganancia resultante de una estrategia activa de inversión cuando se logra acertar el 100% de las pocisiones tomadas.

## Resultados Observados

El modelo construido arrojó una rentabilidad de 170% anualizado sobre data nueva (nunca antes expuesta al modelo), con un promedio de 2,1% de volatilidad negativa.

También se halló que de 96 posiciones tomadas, 66 fueron acertadas, lo cual dio un 65% de aciertos respecto al *perfect equity curve*. El máximo *drawdown*[5] para esta estrategia fue de 5,5 por ciento.

| Resultados de la estrategia propuesta | |
|---|---:|
| Observaciones (Semanas) | 96 |
| Aciertos | 66 |
| Desaciertos | 30,00 |
| Porcentaje de aciertos | 69% |
| Inversión inicial (US$) | 10,00 |
| Capital final (US$) | 62,70 |
| Rendimiento anual | 170% |
| Volatilidad negativa promedio | -2,1% |
| Valor final del perfect equity | 96,26 |
| Cum. Equity / Perfect Equity | 65% |

---

[5] *Drawdown:* es la medida de la caída en el valor de una inversión de un periodo de una serie de tiempo al siguiente.

# Conclusiones

- Las redes neurales ofrecen un modelo viable para hallar relaciones entre datos que difícilmente pueden ser tratados con técnicas estadísticas convencionales.
- El buen desempeño que puedan tener los resultados de la red neural dependen de una cuidadosa selección de las variables de entrada del modelo.
- Una forma para reducir la incertidumbre y escoger adecuadamente las variables de entrada al modelo estriba en el uso apropiado de un sistema dinámico que represente al sistema.
- La inclusión de otros países con características similares a Venezuela mejoró la eficiencia predictiva de la red. La capacidad predictiva del modelo se puede mejorar incluyendo inputs procedentes de otros países relevantes al modelo, ya que es la información del mercado como agregado quien puede influir en los precios del petróleo.
- Dada la construcción de una red que lo refleja, se corrobora la relación que existe entre los gastos nacionales de un país productor y los precios de su petróleo.
- El rendimiento obtenido por un inversionista al seguir la estrategia planteada habría resultado altamente rentable y de bajo riesgo.

# Bibliografía